\documentclass[10pt,aps,twocolumn,nofootinbib]{revtex4-1}
\pdfoutput=1
\usepackage{amsfonts}
\usepackage{mathrsfs}
\usepackage{amsmath}
\usepackage{amssymb}
\usepackage[normalem]{ulem}
\usepackage{extarrows}
\usepackage{slashed}
\usepackage{isodateo}
\usepackage{graphicx}
\usepackage{xcolor}
\usepackage{paralist}
\usepackage[bookmarksnumbered=true,bookmarksopen=true]{hyperref}
 \hypersetup{colorlinks,%
             linkcolor=[rgb]{0,0.3,0.6}, %
             citecolor=[rgb]{0,0.3,0.6}, %
             urlcolor=[rgb]{0,0.3,0.6}}%
\begin{document}

\title{Probing Electroweak Dark Matter at 14 TeV LHC}
\author{Shuai Xu$^{1,}$}
\email{shuaixu@cqu.edu.cn}
\author{Sibo Zheng$^{1,2,}$}
\email{sibozheng.zju@gmail.com}
\affiliation{$^{1}$Department of Physics, Chongqing University, Chongqing 401331, China \\
$^{2}$Department of Physics, Harvard University, Cambridge, MA 02138, USA}

\begin{abstract}
Well-motivated electroweak dark matter is often hosted by an extended electroweak sector 
which also contains new lepton pairs with masses near the weak scale.
In this paper, we explore such electroweak dark matter via combining dark matter direct detections and high-luminosity LHC probes of new lepton pairs.
Using $Z$- and $W$-associated electroweak processes with two or three lepton final states,
we show that dependent on the overall coupling constant, 
dark matter mass up to $170-210$ GeV can be excluded at $2\sigma$ level and up to $175-205$ GeV can be discovered at $5\sigma$ level at the 
14 TeV LHC with integrated luminosities 300 fb$^{-1}$ and 3000 fb$^{-1}$, respectively.
\end{abstract}
\maketitle

\section{Introduction}
Dark Matter (DM) with mass of a few hundred GeVs, 
which is known as the weakly interacting massive particle (WIMP),
is one of leading motivations to the weak-scale new physics beyond standard model (SM).
Any such new physics, if exists, should play an important role at the LHC.
When a WIMP-like DM is in the reach of DM direct-detection experiments, 
the LHC is useful for a double check.
What is even more interesting is that if the WIMP-like DM is beyond the reach of DM direct detections,
the LHC serves as an alternative discovery platform.
This paper is devoted to address the question - what is the potential of high-luminosity (HL) \cite{1310.8361,1812.07831,1902.10229} LHC  probe of general WIMP-like DM.

WIMP-like DM can be hosted in a variety of weak-scale new physics.
For instance, it can be identified as the neutral fermion of the fourth-generation lepton model \cite{9903387,9804359}, 
where there are mixing effects between the new leptons and SM leptons,
the lightest neutralino of the minimal supersymmetric standard model (MSSM) \cite{9709356},
of next-to-minimal supersymmetric standard model (NMSSM) \cite{0910.1785},
or the lightest neutral fermion of singlet-doublet (SD) \cite{0510064,0705.4493,1109.2604} 
and vector-like (VL) lepton model \cite{0910.2732,1706.01071,1711.05362,1904.10145}, 
where there are no direct mixing with SM leptons.
In the various contexts as above,
there are a diversity of interactions and model parameters.
As a result, it seems difficult to work out universal predictions on such WIMP-like DM.
To address this question, the first task is to find a framework viable for most of sophisticated WIMP-like DM models.

The structure of this paper is organized as follows.
In Sec.II, we will employ a framework which can describe three benchmark WIMP-like DM models above,
where model parameters will be introduced.
In Sec.III, we uncover the DM parameter space which satisfies the DM relic density and survives in the latest DM direction-detection limits.
Sec. IV is devoted to explore the exclusion as well as discovery potentials on the DM mass ranges in the light of LHC probes of lepton pairs in the $Z$- and $W$-associated electroweak processes such as $pp \to \ell^\pm \ell^\mp+ E_T^{\rm miss}$,
$pp \to \ell^\pm \ell^\mp jj + E_T^{\rm miss}$ 
and $pp \to \ell^\pm \ell^\mp \ell^\pm + E_T^{\rm miss}$ at the HL-LHC,
where $\ell$, $j$ and $E_T^{\rm miss}$ refer to lepton, jet and missing energy, respectively,
and the samples used for event simulations are directly extracted from the parameter space of electroweak DM. 
Finally, we conclude in Sec.V.

\section{Extended Electroweak Sector}
In various contexts of weak-scale new physics, 
the SM electroweak sector is extended by a couple of electroweak doublets $E^{\pm}$
\begin{eqnarray}{\label{components}}
E^{+}=\left(%
\begin{array}{c}
\eta^{+} \\
\eta^{0}\\
\end{array}%
\right), ~~
E^{-}=\left(%
\begin{array}{c}
\tilde{\eta}^{0} \\
\eta^{-}\\
\end{array}%
\right),
\end{eqnarray}
for which the effective Lagrangian at the weak scale can be described as 
\begin{eqnarray}{\label{Lag}}
\mathcal{L}\supset \frac{i}{2}\overline{E^{+}}\slashed{D}E^{+} + \frac{i}{2}\overline{E^{-}}\slashed{D}E^{-}-m_{E}E^{+}E^{-},
\end{eqnarray}
with $m_E$ a vectorlike (VL) mass.
If there are no other mass sources for the doublets, 
the charged fermion $\chi^{\pm}$ composed of $(\eta^{+},\overline{\eta^{-}})$ 
and the neutral fermion $\chi^{0}$ composed of $(\eta^{0},\tilde{\eta}^{0})$
will have nearly degenerate masses $m_{\chi^{\pm}}\simeq m_{\chi^{0}}$ up to small radiative correction \cite{9804359}.
This mass degeneracy kinematically suppresses the decay $\chi^{\pm}\rightarrow \chi^{0}W^{\pm}$, 
and reduces the signal of the lepton pair $\chi^{+}\chi^{-}$.

The mass degeneracy disappears whenever there are moderate or large mixing effects. 
The mixing effects can be classified into two different types.
In the first type,
$E^{\pm}$ directly mix with SM leptons as in the fourth-generation lepton models.
In the second type, 
$E^{\pm}$ mix with some new fermion singlet $N$ as in the MSSM, NMSSM, SD and VL  lepton models,
which gives rise to sufficient splitting between the charged and neutral fermion masses. 
The scope of this paper is restricted to the large mass splitting driven by the electroweak singlet fermion.

In this case,
the Lagrangian $\mathcal{L}$ in Eq.(\ref{Lag}) is modified by
\begin{eqnarray}{\label{mLag}}
\delta\mathcal{L}\supset \frac{i}{2} N\slashed{\partial}N-\frac{m_{N}}{2}N^{2}-(y_{1}N\overline{E^{+}}+y_{2}NE^{-})H+\text{h.c},\nonumber\\
\end{eqnarray}
where $m_N$ is the singlet mass, 
and $H$ refers to the SM Higgs doublet with a vacuum expectation value (vev) $\upsilon=174$ GeV.
With the new Yukawa interactions in Eq.(\ref{mLag}), 
the neutral fermions are now composed of $(N,\eta^{0}, \tilde{\eta}^{0})$,
whose mass matrix $M_{\chi^{0}}$ reads as,
\begin{eqnarray}{\label{mixing}}
\mathcal{M}_{\chi^{0}}=\left(
\begin{array}{ccc}
 m_{N} & y_{1}\upsilon & y_{2}\upsilon \\
 * & 0  & m_{E} \\
  * &* & 0\\
\end{array}%
\right).
\end{eqnarray}
The mixing effects in Eq.(\ref{mixing}) are responsible for the mass splitting between the three neutral fermion mass $m_{\chi^{0}_{i}}$ ($i=1-3$) and charged fermion mass $m_{\chi^{\pm}}$.

Under the basis of mass eigenvalues,
the Lagrangian is rewritten as
\begin{eqnarray}{\label{eigenstate}}
\mathcal{L}\supset&-&h\overline{\chi^{0}_{i}}\left(c_{h,ij}P_{L}+c^{*}_{h,ij}P_{R}\right)\chi^{0}_{j}\nonumber\\
&-&Z_{\mu}\overline{\chi^{0}_{i}}\gamma^{\mu}\left(c_{Z,ij}P_{L}-c^{*}_{Z,ij}P_{R}\right)\chi^{0}_{j}\nonumber\\
&-&\left[\frac{g}{\sqrt{2}}W_{\mu}^{-}\overline{\chi^{0}_{i}}\gamma^{\mu}(N_{i3}P_{L}-N_{i2}P_{R})\chi^{+}+\text{h.c}\right]
\end{eqnarray}
with
\begin{eqnarray}{\label{coupling}}
c_{Z,ij}&=& \frac{g}{4c_{W}} \left(U_{i3}U^{*}_{j3}-U_{i2}U^{*}_{j2}\right),\nonumber\\
c_{h,ij}&=& \frac{1}{\sqrt{2}}\left(y_{2}U_{i3}U^{*}_{j1}+y_{1}U_{i2}U^{*}_{j1}\right),
\end{eqnarray}
where $c_W$ and $g$ denote the weak mixing angle and weak gauge coupling constant respectively, 
and the unitary matrix $U$ is introduced to diagonalize $\mathcal{M}_{\chi^{0}}$.
We have written fermions in the 4-component notation, 
with $\chi^{+}=(\eta^{+}, \overline{\eta^{-}})$ and $\chi^{0}_{i}=(\chi^{0}_{i}, \overline{\chi^{0}_{i}})$.
In this setting, the lepton pair is a couple of $\chi^{\pm}$ and $\chi^{0}_{i}$ ($i=2,3$), 
with their couplings to the SM sector given in Eq.(\ref{eigenstate}).

\section{Parameter Space of Electroweak DM}
With the lightest neutral state $\chi^{0}_{1}$ identified as the thermal electroweak DM, 
we can work out the parameter space of DM relic density from Eq.(\ref{mixing}) and Eq.(\ref{eigenstate}).
In the framework above, there are four parameters $y_1$, $y_2$, $m_N$ and $m_E$.
Since electroweak DM favors coupling $y=\sqrt{y^{2}_{1}+y^{2}_{2}}$ of order weak interaction $\sim 0.3-1.0$ and DM mass of order 
weak scale $\sim 100-1000$ GeV, we consider the following parameter regions, 
\begin{eqnarray}{\label{ranges}}
 0.3\leq &y&\leq 1.2,\nonumber\\
100~\rm{GeV} \leq &m_{E}& \leq 1500~\rm{GeV},\\
100~\rm{GeV} \leq &m_{N}& \leq 1500~\rm{GeV},\nonumber
\end{eqnarray}
over which we perform random scans, with $\rm{sign}(y_{2}/y_{1})$ either positive or negative.

\begin{table}
\begin{center}
\begin{tabular}{cccc}
\hline\hline
 $\rm{model}$ & ~~$\rm{fields}$~~  & $y$  & $\rm{sign}(y_{2}/y_{1})$ \\ \hline
$\rm{MSSM}$ & ~~$\tilde{B}^{0}$, $\tilde{h}_{u}^{0}$, $\tilde{h}_{d}^{0}$~~ &  $0.35$ & $-$ \\
~~  & ~~$\tilde{W}^{0}$, $\tilde{h}_{u}^{0}$, $\tilde{h}_{d}^{0}$~~ &  $0.63$ & $-$ \\
\hline
$\rm{SD}$ & ~~$N$, $E^{0}$, $\tilde{E}^{0}$~~ &  $0.7-1.0$ & $\pm$ \\
$\rm{VL}$ & ~~$N$, $E^{0}$, $\tilde{E}^{0}$~~ &  $1.0-1.2$ & $\pm$ \\
\hline \hline
\end{tabular}
\caption{The matter content, the magnitude of $y$, 
and $\rm{sign}(y_{2}/y_{1})$ in various benchmark WIMP-like DM models.
We focus on the parameter regions in the SD model which differ from the MSSM, 
and those in the VL model which are favored by the SM Higgs mass constraint.}
\label{benchmark}
\end{center}
\end{table}

\begin{figure*}
\centering
\includegraphics[width=8cm,height=7cm]{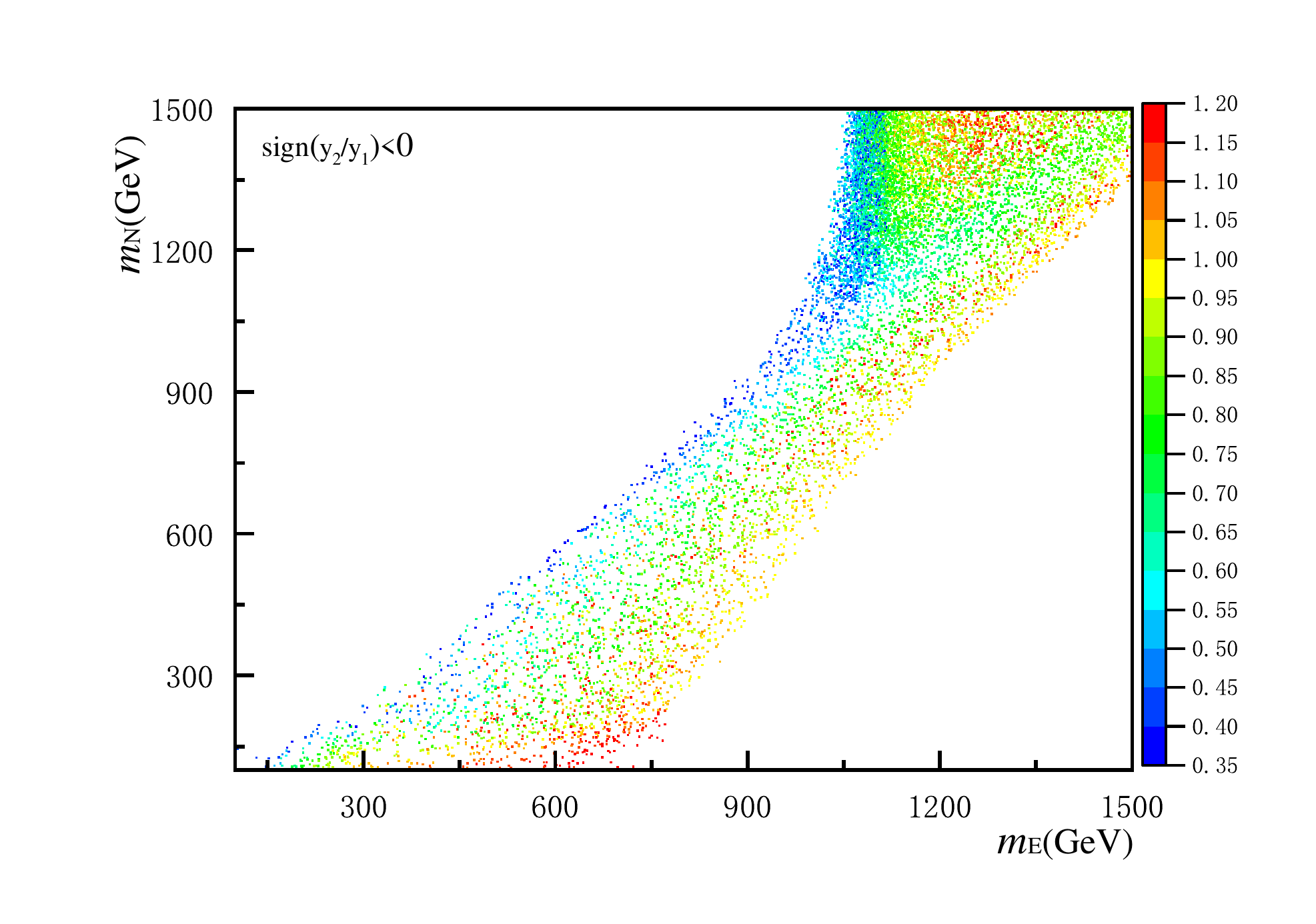}~~~~
\includegraphics[width=8cm,height=7cm]{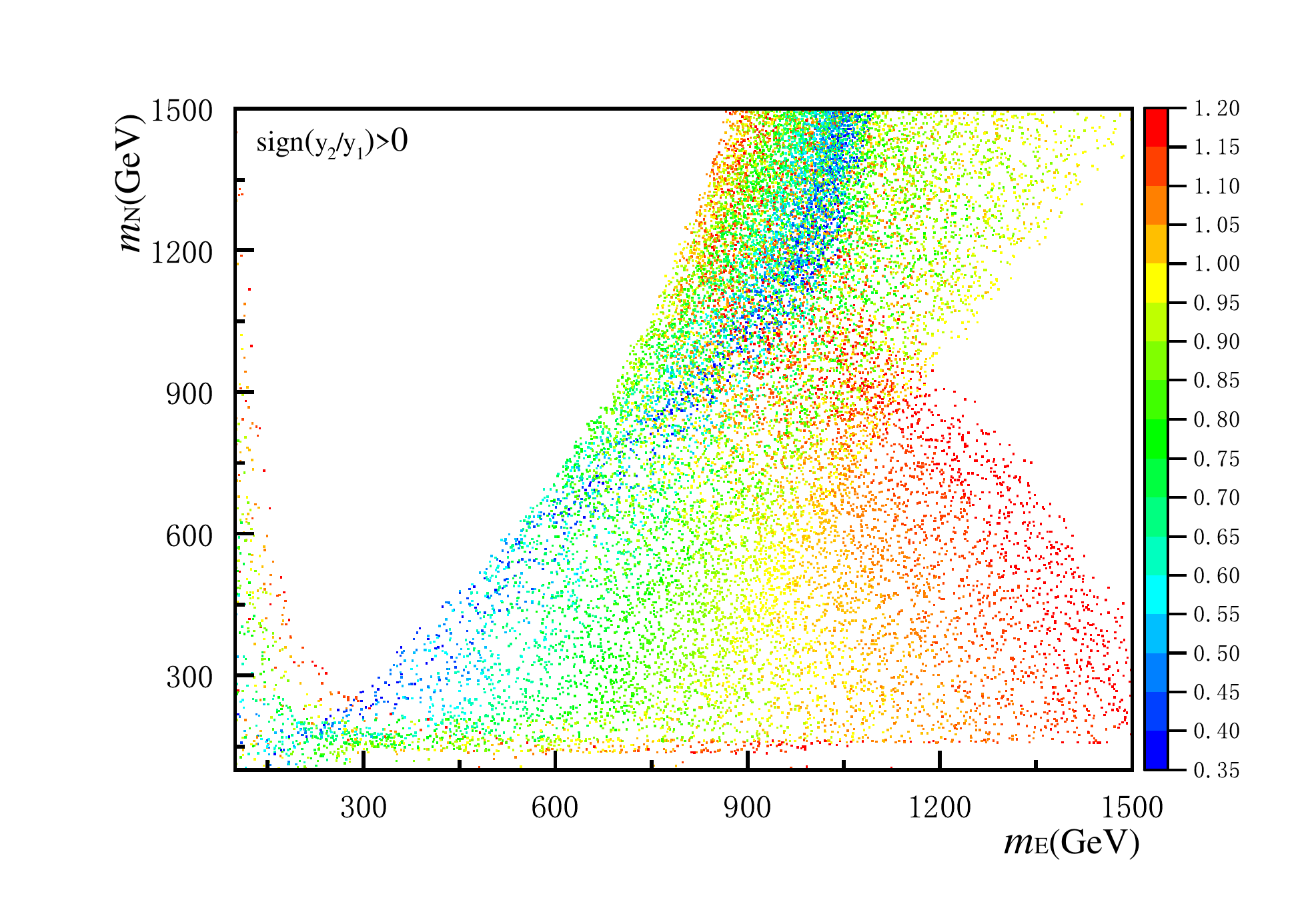}\\
\includegraphics[width=8cm,height=7cm]{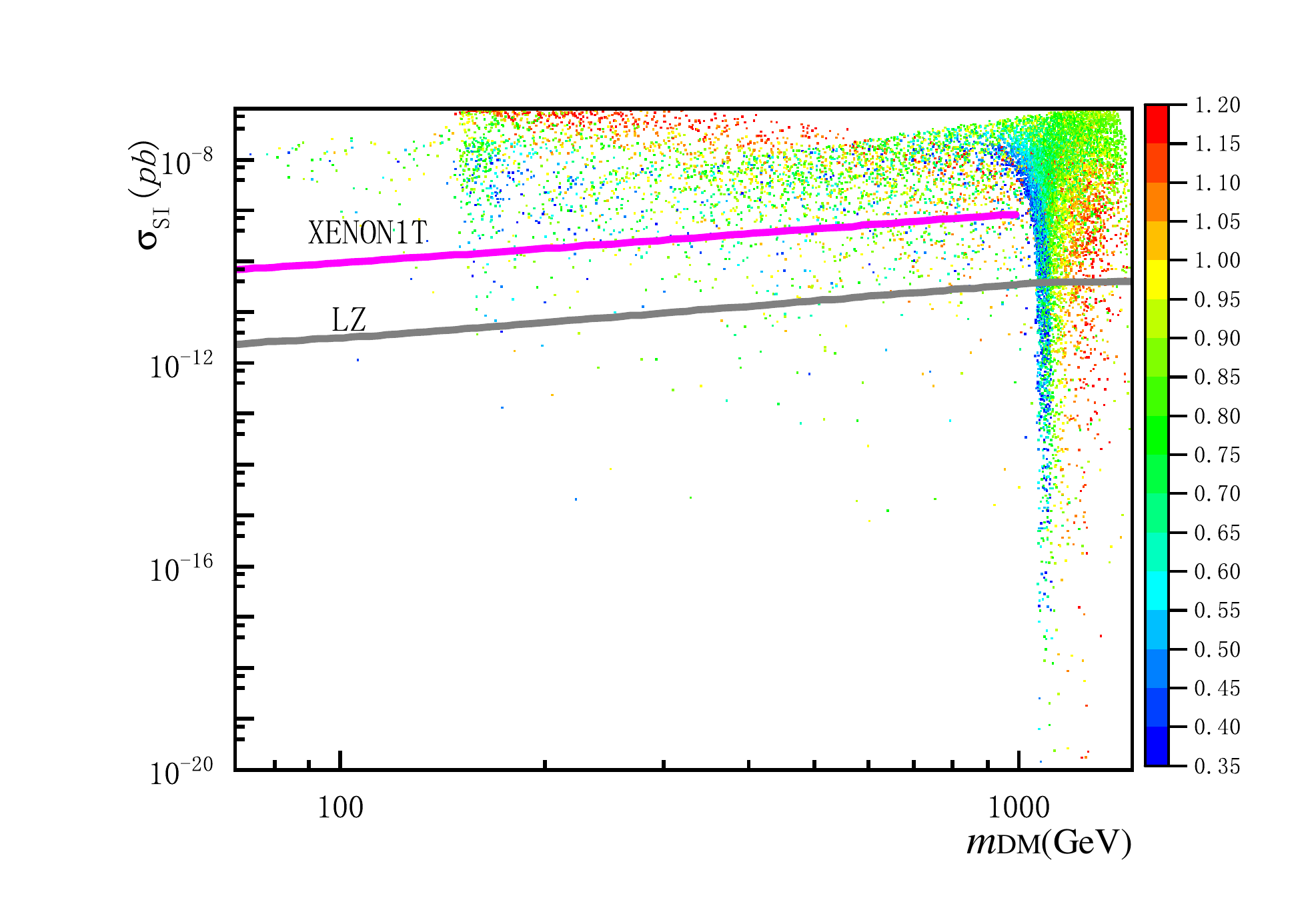}~~~~
\includegraphics[width=8cm,height=7cm]{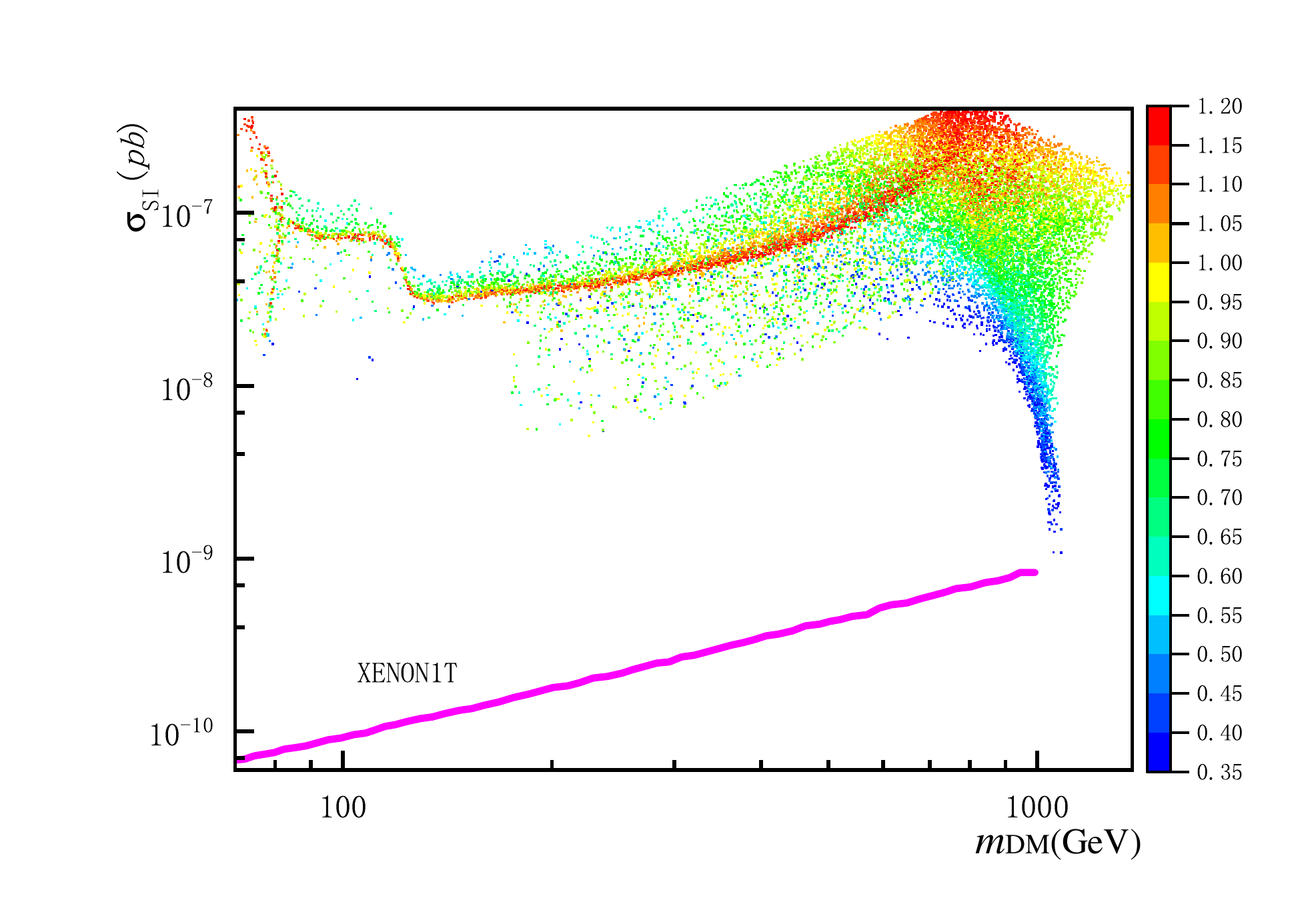}\\
\includegraphics[width=8cm,height=7cm]{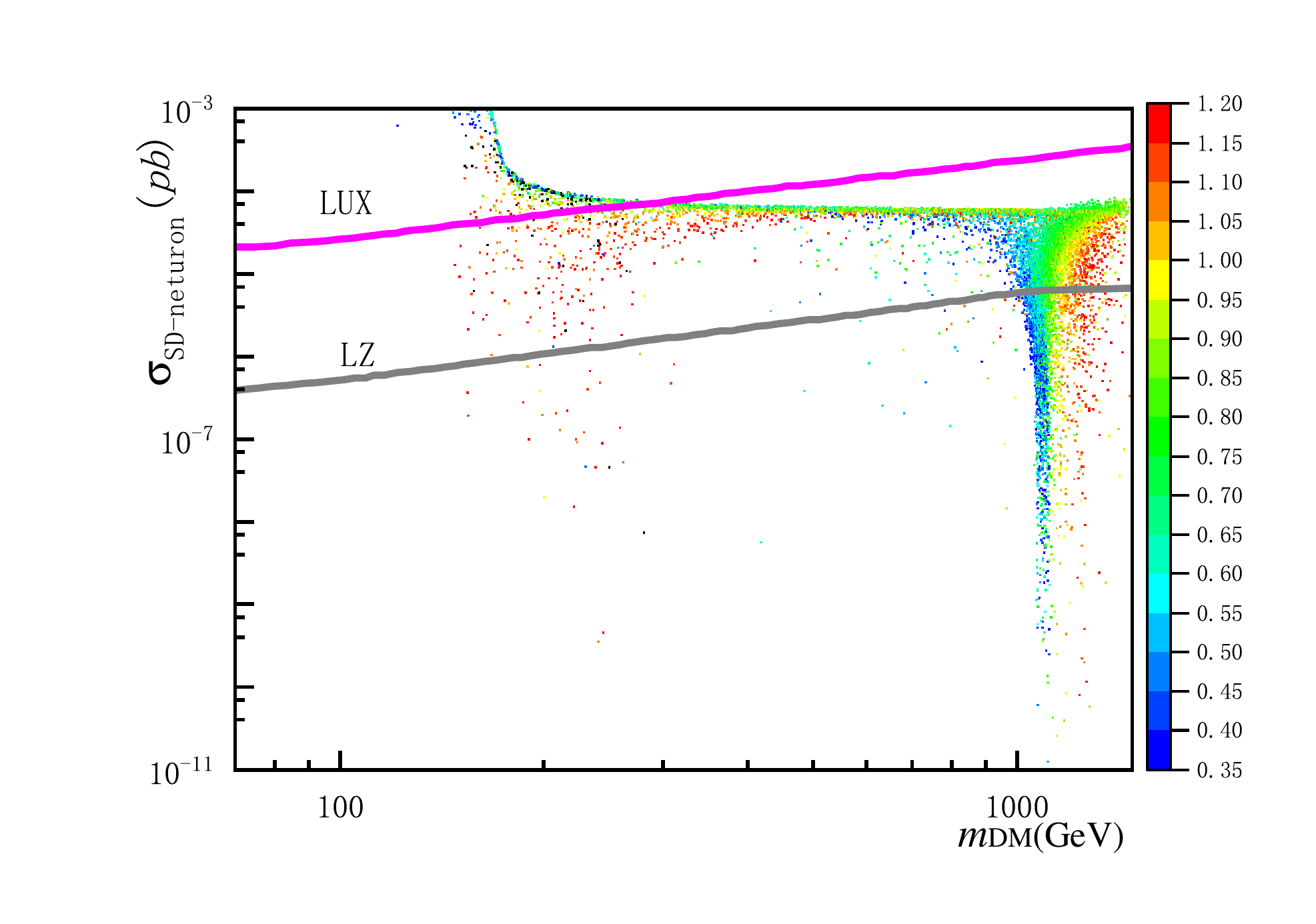}~~~~
\includegraphics[width=8cm,height=7cm]{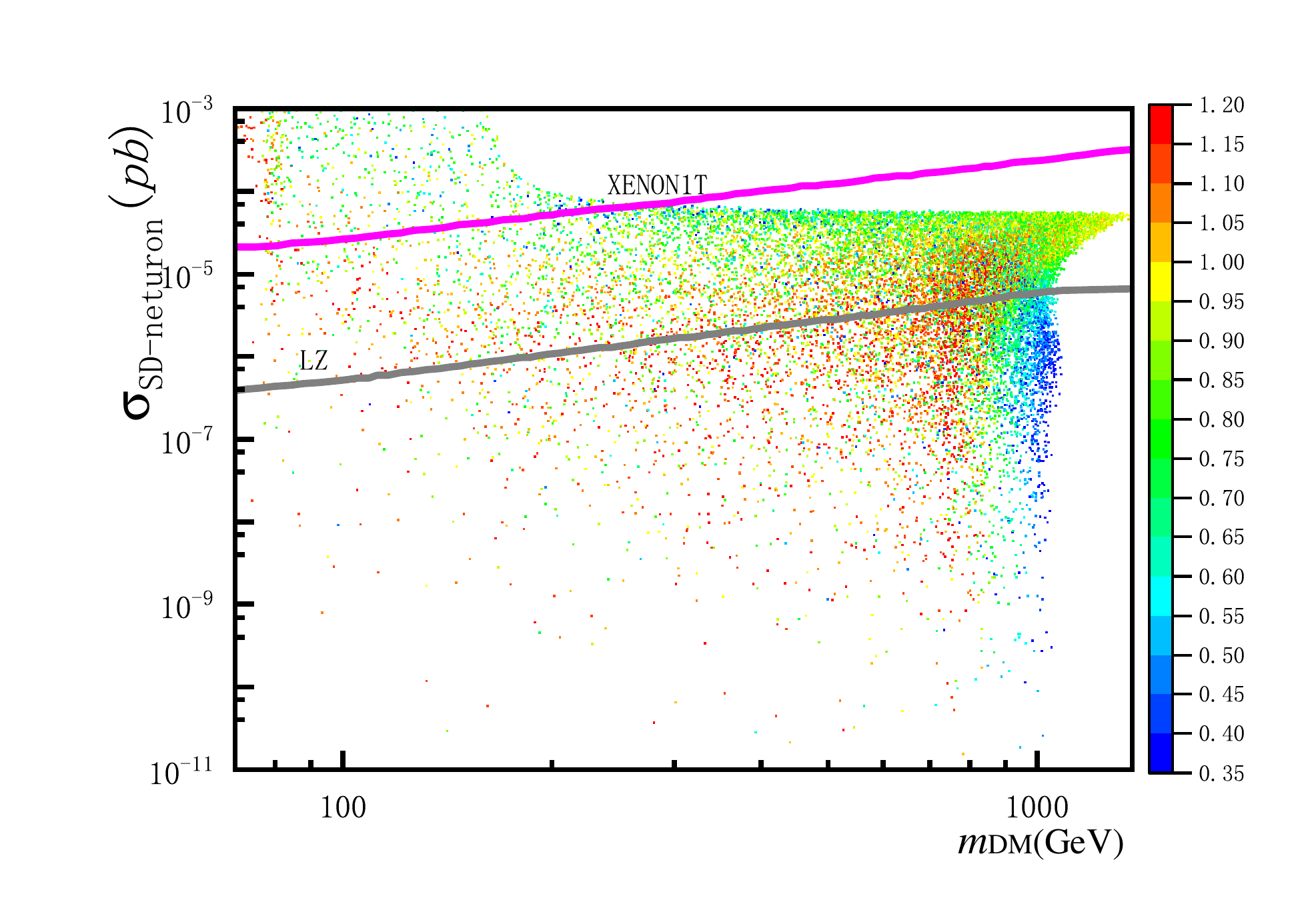}
\caption{The \emph{left} and \emph{right} plots correspond to $\text{sign}(y_{1}/y_{2})<0$ and $\text{sign}(y_{1}/y_{2})>0$, respectively, 
where we show the parameter space of DM relic density $\Omega\rm{h}^{2}=0.12\pm 0.005$ (\rm{top}), 
the SI $\chi^{0}_{1}$-nucleon scattering cross section (\rm{middle}) 
and the SD $\chi^{0}_{1}$-neutron scattering cross section (\rm{bottom}) as function of DM mass $m_{\text{DM}}$ and coupling $y$, 
with the Xenon1T \cite{Xenon1T}, LUX \cite{LUXSD} and the future LZ \cite{LZ} limits shown for comparison.}
\label{ps}
\end{figure*}

We employ the code micrOMEGAs \cite{1407.6129} to calculate the relic density of the lightest neutral fermion $\chi^{0}_{1}$,
and record the spin-independent  (SI) and spin-dependent (SD) $\chi^{0}_{1}$-nucleon scattering cross sections.
Shown in Fig.\ref{ps} are the parameter spaces of the electroweak DM with $\text{sign}(y_{2}/y_{1})<0$ (\emph{left}) and $\text{sign}(y_{2}/y_{1})>0$ (\emph{right}), respectively.
The two types of plots therein clearly indicate that the DM relic density is sensitive to both the magnitude of $y$ and $\rm{sign}(y_{2}/y_{1})$. 
We refer the reader to Table.\ref{benchmark} about $\rm{sign}(y_{2}/y_{1})$ in the benchmark electroweak DM models.

A few comments are in order regarding the parameter space of electroweak DM.
Firstly,  the left plots with $\rm{sign}(y_{2}/y_{1})<0$  cover simplified MSSM\cite{1211.4873,1306.1567,1412.5952,1506.07177,1701.05869}, SD \cite{1311.5896,1505.03867} 
and VL \cite{1904.10145} models. 
\begin{itemize}
\item  For the simplified MSSM, the coupling $y=e/\cos\theta_{W}$ and $y=e/\sin\theta_{W}$ for simplified bino-higgsino and wino-higgsino systems, respectively, with $\theta_{W}$  the weak mixing angle. 
In both simplified models, $y_{2}/y_{1}\rightarrow -\tan\beta$, where $\tan\beta$ is the ratio of two vacuum expectation values (vevs) of Higgs doublets.  The left middle and bottom plots show that a portion of samples with $y=0.35$ \cite{1701.05869} or $y=0.63$ \cite{1506.07177} survives.
\item  For either SD or VL model, the parameter space with coupling $y>1$ is nearly excluded for DM mass beneath $1$ TeV if not all,
in the light of the Xenon1T limit as shown in the left middle plot.
This plot suggests that both the SD and VL models tend to survive in the small $y$ region,
which is consistent with previous observations on the SD model in \cite{1311.5896,1505.03867} and the VL model in \cite{1904.10145}, respectively. 
\end{itemize}

Secondly, the right plots with $\rm{sign}(y_{2}/y_{1})>0$ cover the SD \cite{1311.5896,1505.03867} and VL \cite{1904.10145} models. 
\begin{itemize}
\item Comparing the two middle plots in Fig.\ref{ps} shows that 
the Xenon1T limit is more stringent in the situation with $\rm{sign}(y_{2}/y_{1})>0$,
although the number of samples is relatively larger in this case.
It turns out that DM mass between $1$ TeV is almost excluded by the Xenon1T limit in this case,
which strengthens the earlier results in ref.\cite{1505.03867}.
\item The importance of $\rm{sign}(y_{2}/y_{1})$ on the SI or SD cross section can be understood
from earlier analysis on the blind spots in ref.\cite{1211.4873},
under which either the SI or SD cross section is dramatically suppressed.
From our setup in Eq.(\ref{ranges}), the condition of blind spots for the SI cross section favors a negative $\rm{sign}(y_{2}/y_{1})$.
\end{itemize}

Finally, we would like to mention that our analysis not only reproduces,
but also expands the parameter space beyond the three benchmark electroweak DM models as shown in Table.\ref{benchmark}. 
Some of the new parameter regions may be useful for other models of new physics around the weak scale.
Moreover, one observes that a portion of samples is even below the sensitivity of future LZ experiment.
However, they may be in the reach of future LHC, which is the subject as what follows.

\begin{table}
\begin{center}
\begin{tabular}{cccc}
\hline\hline
${\rm lepton~pair}$&${\rm SM~background}$ &  ~~${\rm refs}$   \\  \hline
$\chi^{+}\chi^{-}$& $pp \to \ell_{\alpha/\beta}^\pm \ell_{\alpha/\beta}^\mp+ E_T^{\rm miss}$  &~~ \cite{1403.5294, 2019008,1908.08215} \\ \hline
$\chi^{\pm}\chi^{0}_{2}$&~~ $pp \to \ell_{\alpha/\beta}^\pm \ell_{\alpha/\beta}^\mp jj' + E_T^{\rm miss}$  &~~ \cite{1403.5294,1803.02762,1806.02293} \\
 &~~ $pp \to \ell_\alpha^\pm \ell_\alpha^\mp \ell_{\alpha/\beta}^\pm jj+ E_T^{\rm miss}$  &~~ \cite{1403.5294,1803.02762,1806.02293} \\\hline
$\chi^{\pm}\chi^{0}_{2}$ & $pp \to \ell_\alpha^\pm \ell_\alpha^\mp \ell_{\alpha/\beta}^\pm+ E_T^{\rm miss}$  &~~ \cite{1803.02762,1806.02293} \\ 
\hline \hline
\end{tabular}
\caption{SM backgrounds for the electroweak productions of lepton pairs $\chi^{+}\chi^{-}$ and $\chi^{\pm}\chi^{0}_{2}$,
 with subsequent electroweak decays to $\chi^{0}_{1}$,  where $j, j'$ and $\alpha,\, \beta$ refer to jets and charged leptons $\ell=\{e,\mu\}$.
For example, in the first case the two leptons can have either the same or different flavors with opposite signs.}
\label{channels}
\end{center}
\end{table}

\section{Lepton Pairs at the LHC}
\begin{figure}
\centering
\includegraphics[width=2.5cm,height=3cm]{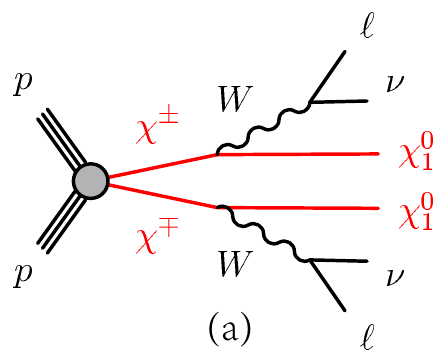}~~
\includegraphics[width=2.5cm,height=3cm]{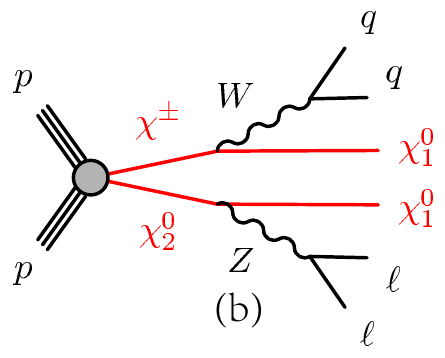}~~
\includegraphics[width=2.5cm,height=3cm]{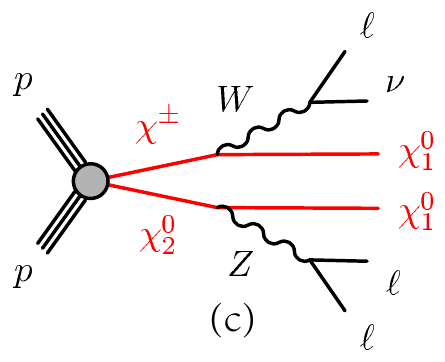}
\caption{Feynman diagrams for the signal channels of new lepton pairs in Table \ref{channels}.}
\label{Feyn}
\end{figure}

\begin{figure}[htb!]
\centering
\includegraphics[width=7cm,height=6cm]{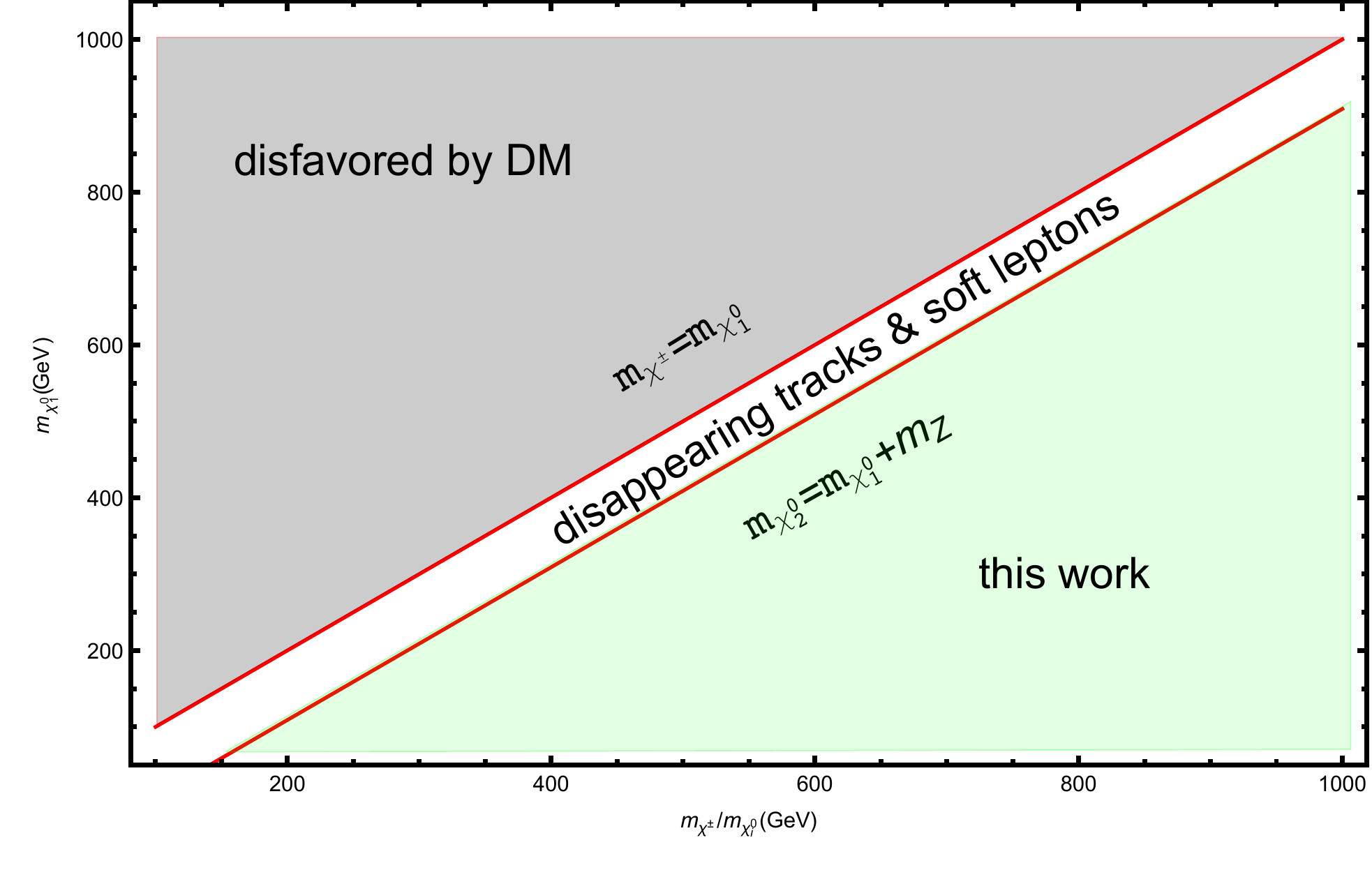}
\caption{An illustration of samples extracted from the parameter space of electroweak DM in the left plots of Fig.\ref{ps} in this work.}
\label{sketch}
\end{figure}

According to the effective Lagrangian in Eqs.(\ref{Lag})-(\ref{mLag}), 
the productions of new lepton pairs at the LHC as shown in Tabel \ref{channels},
for which the Feynman diagrams can be found in Fig.\ref{Feyn},
mainly depend on the electroweak interactions with SM gauge bosons and Higgs scalar.
Here, we focus on the electroweak productions of lepton pairs, 
which subsequently decays into on-shell $Z$ and/or $W$.
The direct combination of DM direct detections and LHC probes of lepton pairs makes our analysis intuitive, 
which differs from either $i)$ the experimental analysis \cite{1403.5294, 1803.02762,1806.02293, 2019008,1908.08215} on the electroweakinos reported by ATLAS and CMS collaborations, 
where the samples were not extracted from the electroweak DM parameter space, 
or $ii)$ the DM direct-detection predictions on the electroweak DM as above, 
where the collider probes are not clear.
Since we focus on the parameter space of electroweak DM as shown in Fig.\ref{sketch}, 
where the final lepton states are standard rather than soft due to large mass splitting between the charged/heavier neutral states and DM, 
our analysis is also different from $iii)$ the studies on the electroweak DM in the parameter space with small mass splitting such as the disappearing tracks \cite{1804.07321,1712.02118,1703.09675,1805.00015}, mono-jets \cite{1805.00015,CMS12048,1312.7350,1311.7641,1506.02148} 
and soft lepton final states \cite{1310.3675, 1307.5952,1404.0682,0909.4549}.

Shown in Table \ref{channels} are the SM backgrounds referring to the electroweak productions of lepton pairs $\chi^{+}\chi^{-}$ and $\chi^{\pm}\chi^{0}_{2}$ at the LHC.
\begin{itemize}
\item  The electroweak production of lepton pair $\chi^{+}\chi^{-}$ is dominated by $pp\rightarrow Z^{*} \rightarrow \chi^{+}\chi^{-}$,
which decays as $\chi^{\pm}\rightarrow W^{\pm}\chi^{0}_{1}$, with $W\rightarrow \ell\nu$.
The SM background for this process is composed of two charged leptons with missing energy $E_T^{\rm miss}$.
\item  The electroweak production of lepton pair $\chi^{\pm}\chi^{0}_{2}$ is similar to that of $\chi^{+}\chi^{-}$,
which decay as $\chi^{\pm}\rightarrow W^{\pm}\chi^{0}_{1}\rightarrow jj\chi^{0}_{1}$ or
$\chi^{\pm}\rightarrow W^{\pm}\chi^{0}_{1}\rightarrow \ell \nu\chi^{0}_{1}$ and $\chi^{0}_{2}\rightarrow Z\chi^{0}_{1}\rightarrow \ell\ell \chi^{0}_{1}$, respectively.
The SM backgrounds are composed of either two leptons and two jets with $E_T^{\rm miss}$ or three leptons with $E_T^{\rm miss}$.
\end{itemize}

For numerical calculation,
we firstly use the package FeynRules \cite{1310.1921} to prepare the model files for MadGraph5 \cite{1405.0301},
which contains package Pythia 8 \cite{Sjostrand:2014zea} for parton showering and hadronization,
and package Delphes 3 \cite{1307.6346} for fast detector simulation.
Then, we use Madgraph5 to generate $2000$ events for each sample 
which is directly extracted from the parameter space of electroweak DM. 
The total number of samples from the left plots in Fig.\ref{ps} is about $\sim 3660$, 
some of which can yield the on-shell decays $\chi^{\pm}\rightarrow \chi^{0}_{1}W^{\pm}$ and/or $\chi^{0}_{2}\rightarrow \chi^{0}_{1}Z$,
and finally reveal both the exclusion and discovery limits at the 14 TeV LHC with the integrated luminosities 300 fb$^{-1}$ and 3000 fb$^{-1}$ respectively.

\begin{table}
\begin{center}
\begin{tabular}{ccccc}
\hline\hline
&$\chi^{+}\chi^{-}$ &~~$\chi^{\pm}\chi^{0}_{2}$ (2$\ell$ 2$j$) &~~  $\chi^{\pm}\chi^{0}_{2}$ (3$\ell$) \\
\hline
$p_{T}^{l_{1(2)}}$  & $> 25$ & $> 25$  & $> 25$ \\
$\eta_{e(\mu)}$   &  $> 2.47(2.7)$ & $> 2.47(2.7)$  &$> 2.47(2.7)$ \\
$m_{l_{1}l_{2}}$  &$> 100(121.2) $  & 81-101   & 81.2-101.2 \\
$E_T^{\rm miss}$  &  $> 110$ & $> 100$  & $> 170$ \\
 $m_{T}$ &$> 100$   & -   & $> 110$
 \\
$n_{\text{non-b}}$ &0  & 2  & 0 \\
  $n_{\text{b}}$ &  0 & 0  &0 \\
$m_{jj}$  &  - & 70-90  &- \\
$E_{T,s}^{\rm miss}$   &  $> 10$ & -  & - \\
\hline \hline
\end{tabular}
\caption{The main cuts used for event selections of both signals and backgrounds as shown in Table.\ref{channels},
where mass parameters are in unit of GeV. See text for definitions and explanations about these cuts.}
\label{cuts}
\end{center}
\end{table}

For selections of events,
we use the 13-TeV cuts reported by the ATLAS Collaboration
in ref.\cite{1803.02762} and ref.\cite{1908.08215} for lepton pair $\chi^{\pm}\chi^{0}_{2}$
with SM backgrounds $pp \to 2 \ell 2j + E_T^{\rm miss}$ or $pp \to 3\ell + E_T^{\rm miss}$ and  lepton pair $\chi^{+}\chi^{-}$ with SM background $pp \to 2\ell  + E_T^{\rm miss}$, respectively.
The effects on the event numbers due to  deviation from the 14-TeV cuts are expected to be in percent level.
We summarize the details on the cuts in Table.\ref{cuts},
where ${p_{T}}^{l_{1(2)}}$ is the transverse momentum of the first (second) leading lepton $\ell=\{e,\mu\}$,
$\eta_{e(\mu)}$ is pseudo-rapidity of $e$($\mu$), 
$m_{l_{1}l_{2}}$ is the invariant mass of the two leptons with same (or different) flavor for the lepton pair $\chi^{+}\chi^{-}$ or the shell mass of same lepton flavor for the lepton pair $\chi^{\pm}\chi^{0}_{2}$.
In this Table, the transverse mass is defined as $m_\textup{T}=\sqrt{2\times|\mathbf p_{\mathrm{T,1}}|\times|\mathbf p_{\mathrm{T,2}}|\times(1-\mathrm{cos}(\Delta\phi))}$, where $\Delta\phi$ is the difference in azimuthal angle between the particles with transverse momenta $\mathbf p_{\mathrm{T,1}}$ and $\mathbf p_{\mathrm{T,2}}$, 
$n_{\rm{b}}$ is the number of b-tagging jets, 
variable $n_{\text{non-b}}$ refers to the number of jets with $p_{T}>30$ GeV that do not satisfy the $b$-tagging criteria,
$m_{jj}$ is the invariant mass of the two leading jets,
and significance of missing energy $E_{T, s}^{\rm miss}$ is linked to the $E_T^{\rm miss}$ by $E_T^{\rm miss}/\sqrt{H}$, with $H$ the transverse momenta of all final states. 

For $pp \to \ell_\alpha^\pm \ell_\beta^\pm jj + E_T^{\rm miss}$ in the second column in Table.\ref{cuts},
we need additional cuts as follows: $\Delta\phi_{(p_{T}^{miss} ,Z)}<0.8$, $\Delta\phi_{(p_{T}^{miss} ,W)}>1.5$, $0.6<E_T^{\rm miss}/p^{Z}_{\rm T}<1.6$ and $E_T^{\rm miss}/p^{W}_{\rm T}<0.8$,
in which the symbols $W$ and $Z$ correspond to the reconstructed $W$ and $Z$ bosons in the final state. Because the $Z$ boson is always reconstructed from the two leptons, whereas the $W$ boson is reconstructed from the two jets. Here, $p^{Z}_{\rm T}$ and $p^{W}_{\rm T}$ are the transverse momenta of dileptons and dijets respectively, with $\Delta\phi_{(p_{T}^{miss} ,Z)}$ or $\Delta\phi_{(p_{T}^{miss} ,W)}$  referring to the azimuthal angle between the dilepton or dijets transverse momentum and
$p_{T}^{miss}$.

\begin{figure*}
\centering
\includegraphics[width=8cm,height=7cm]{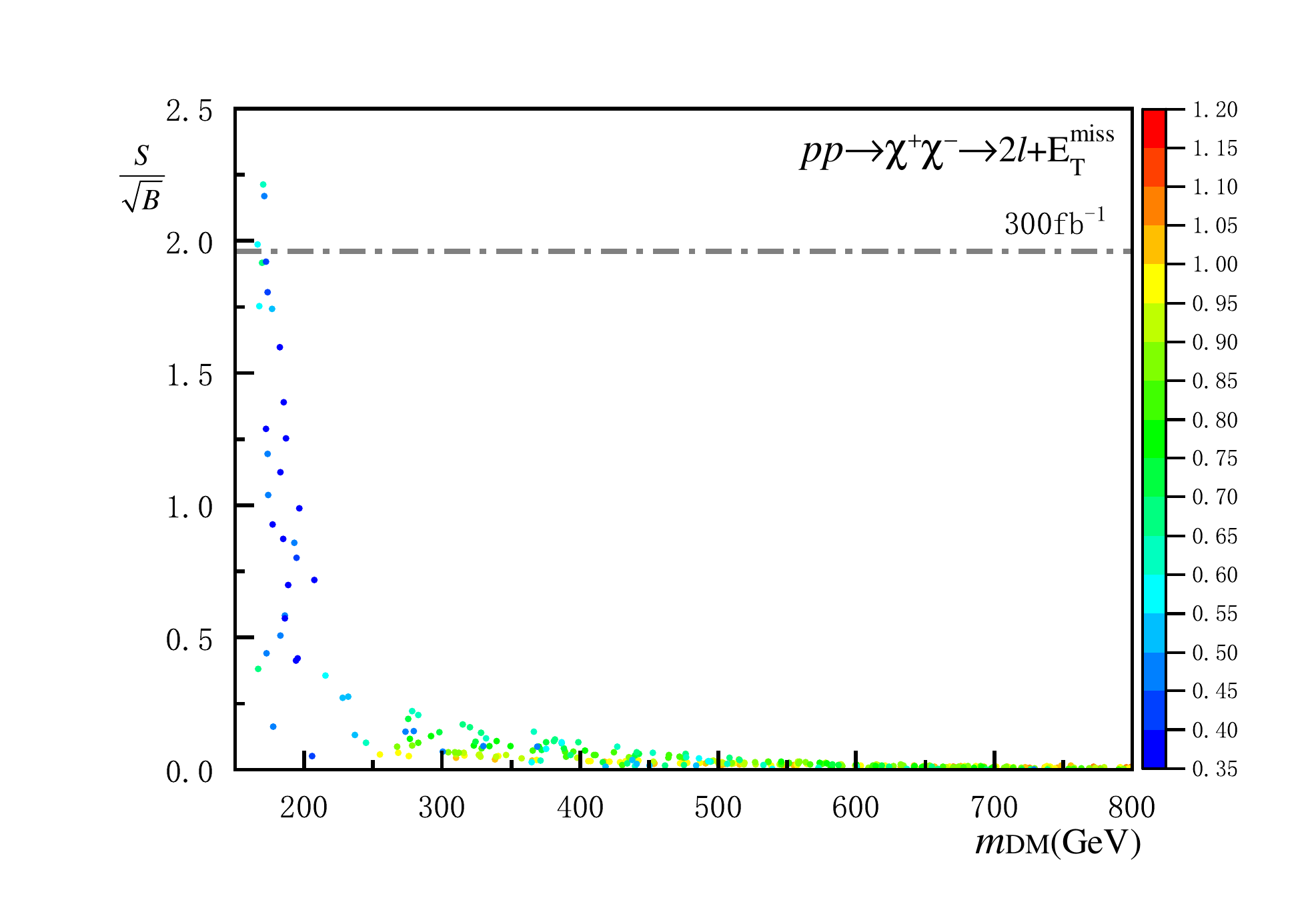}~~~~
\includegraphics[width=8cm,height=7cm]{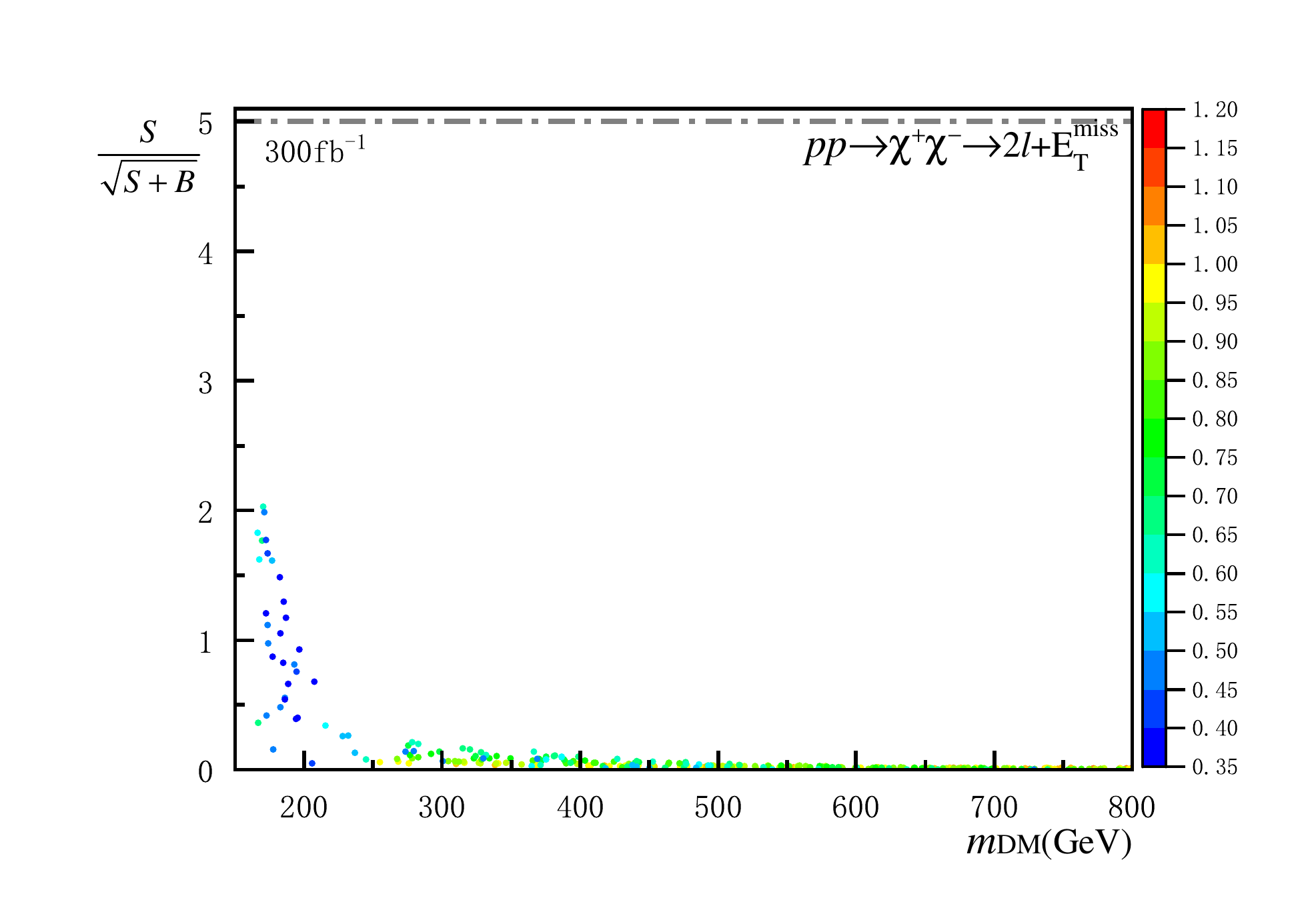}\\
\includegraphics[width=8cm,height=7cm]{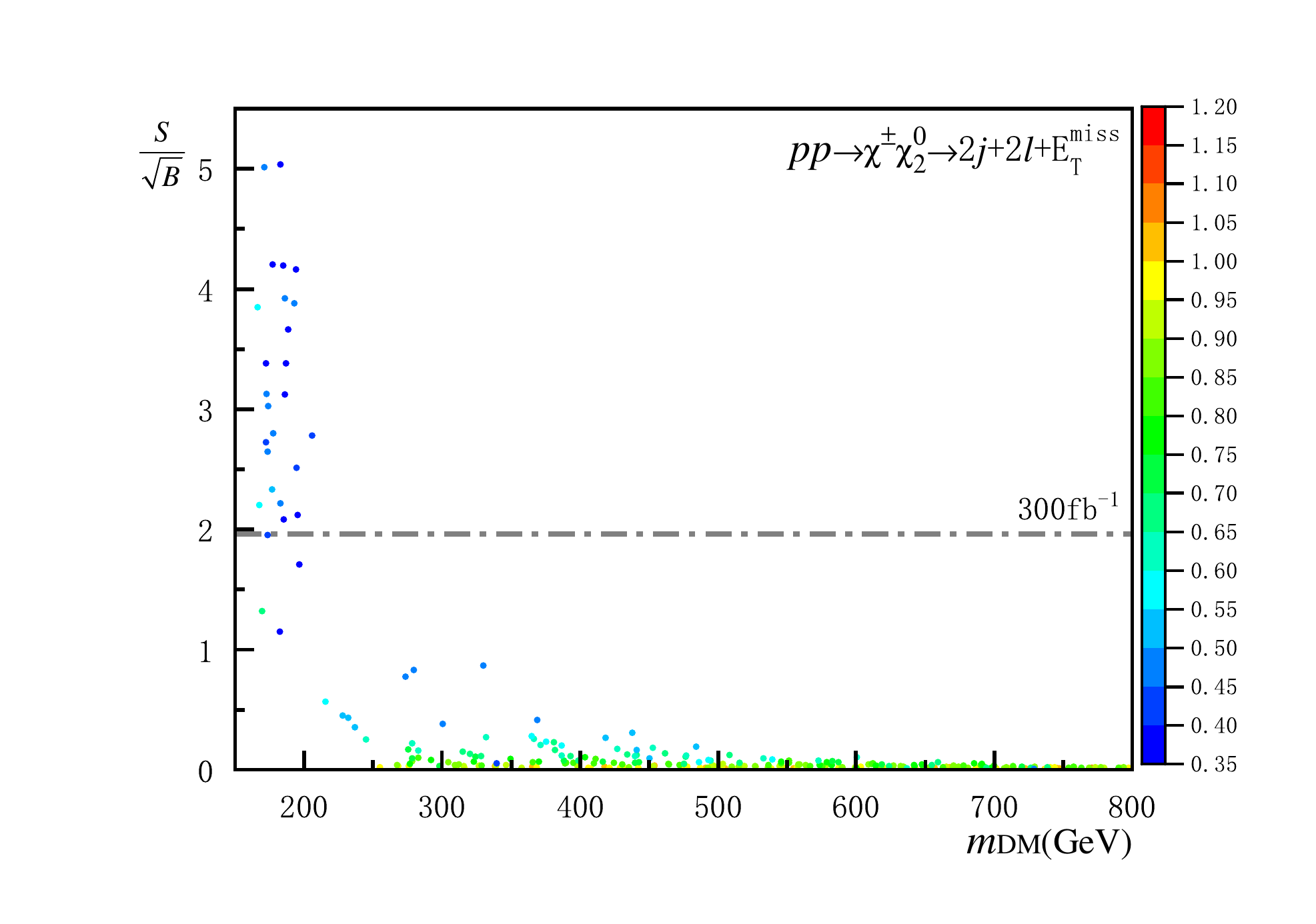}~~~~
\includegraphics[width=8cm,height=7cm]{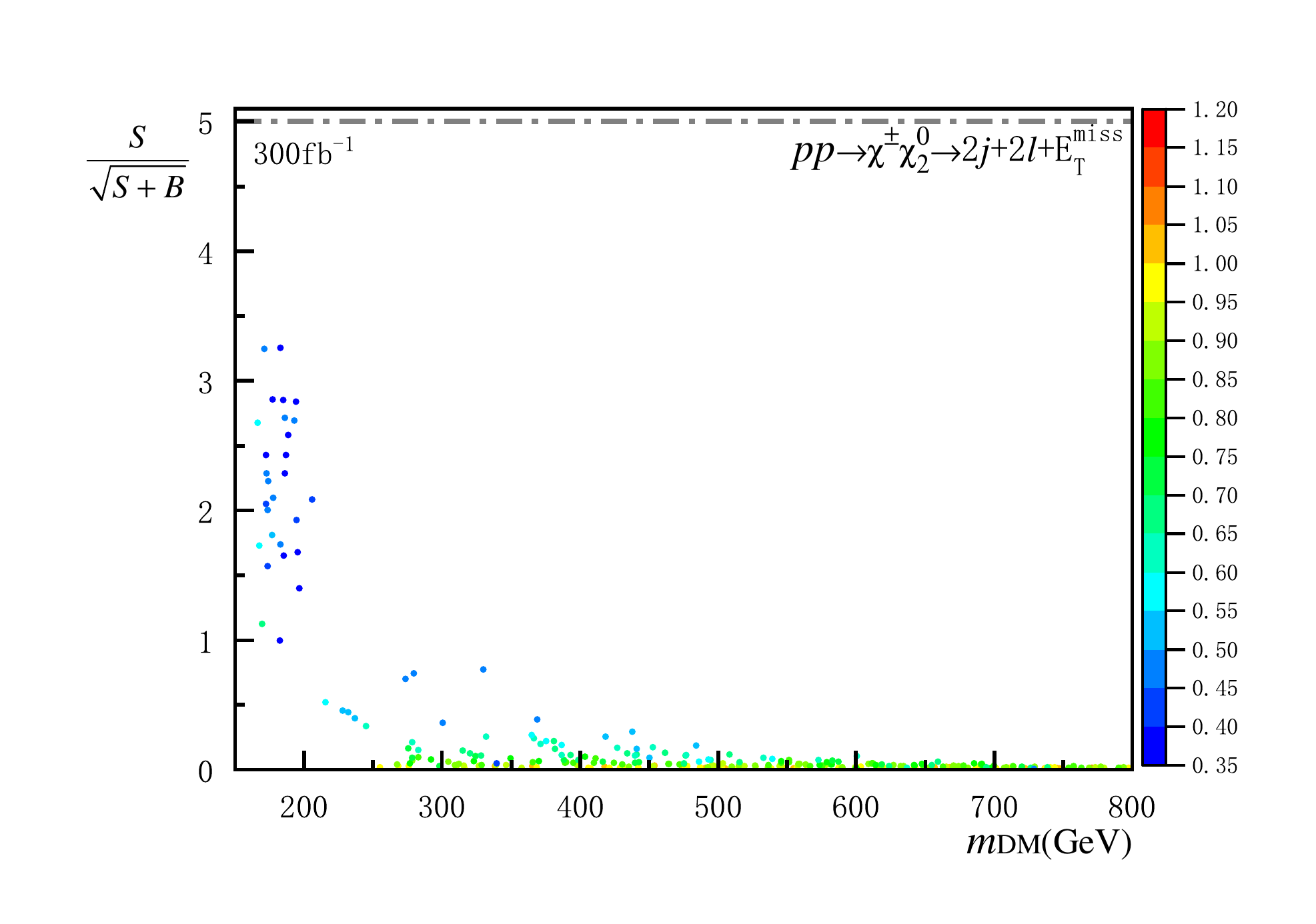}\\
\includegraphics[width=8cm,height=7cm]{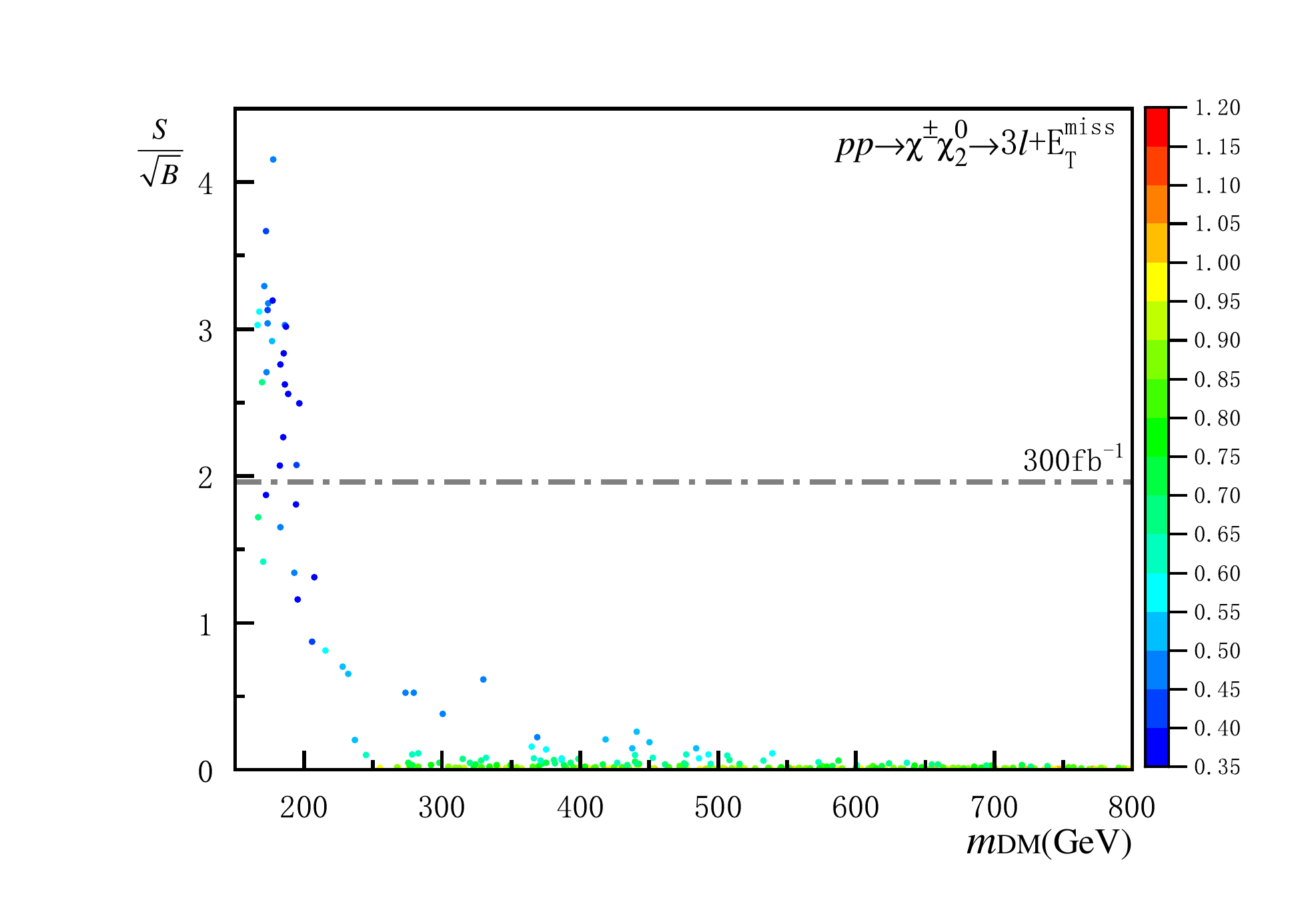}~~~~
\includegraphics[width=8cm,height=7cm]{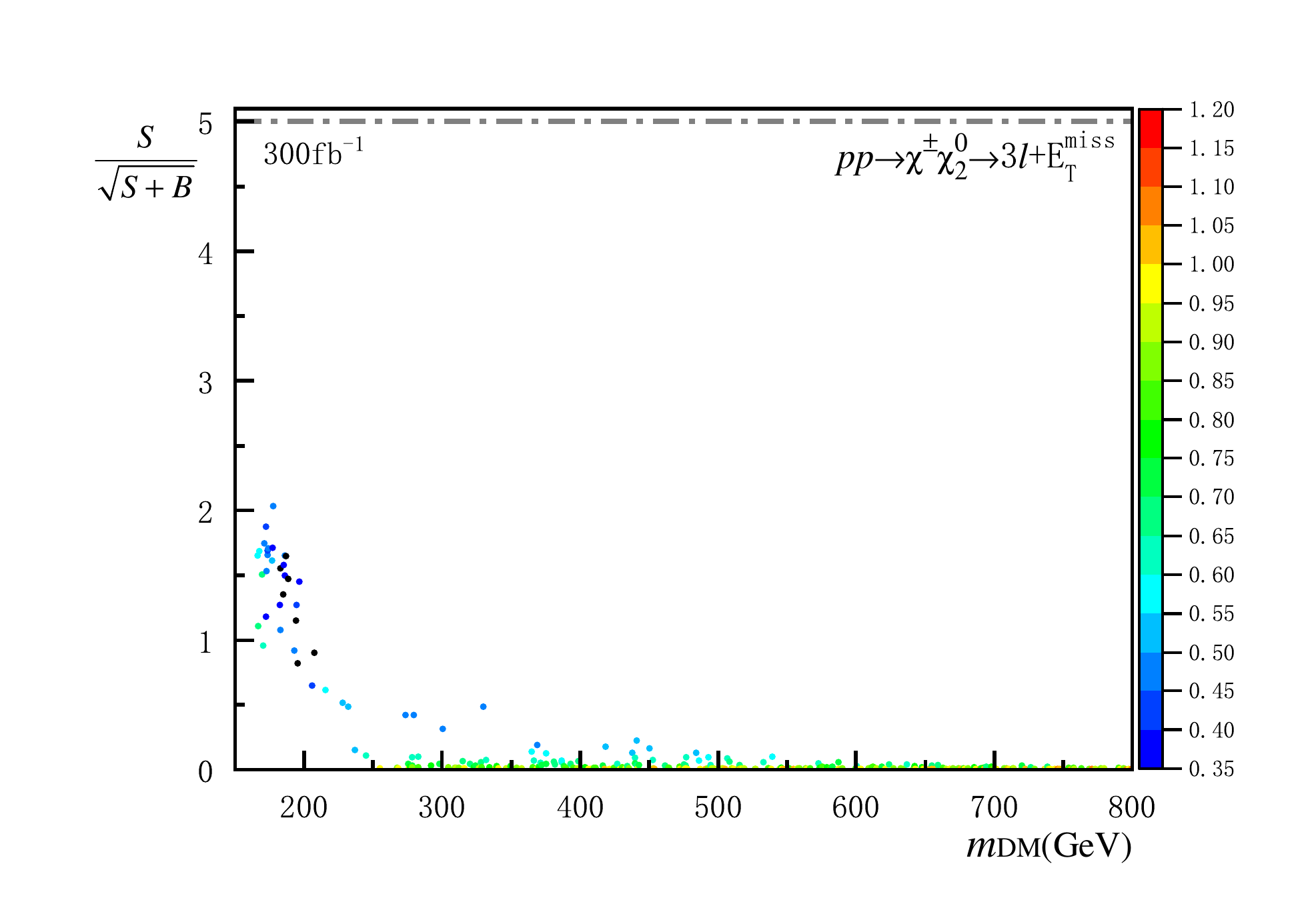}
\caption{The \emph{left} and \emph{right} plots refer to 2$\sigma$ exclusion  and  5$\sigma$ discovery at the 14 TeV LHC with the integrated luminosity $L=300$ fb$^{-1}$,
where the lepton pair $\chi^{+}\chi^{-}$ and $\chi^{\pm}\chi^{0}_{2}$ are shown in the top as well as middle and bottom plots, respectively. 
 All the samples used for event simulations are extracted from the DM parameter space of the \emph{left} plots in Fig.\ref{ps},
which satisfy the DM relic density and survive in the DM direct detections simultaneously.}
\label{potential1}
\end{figure*}

\begin{figure*}
\centering
\includegraphics[width=8cm,height=7cm]{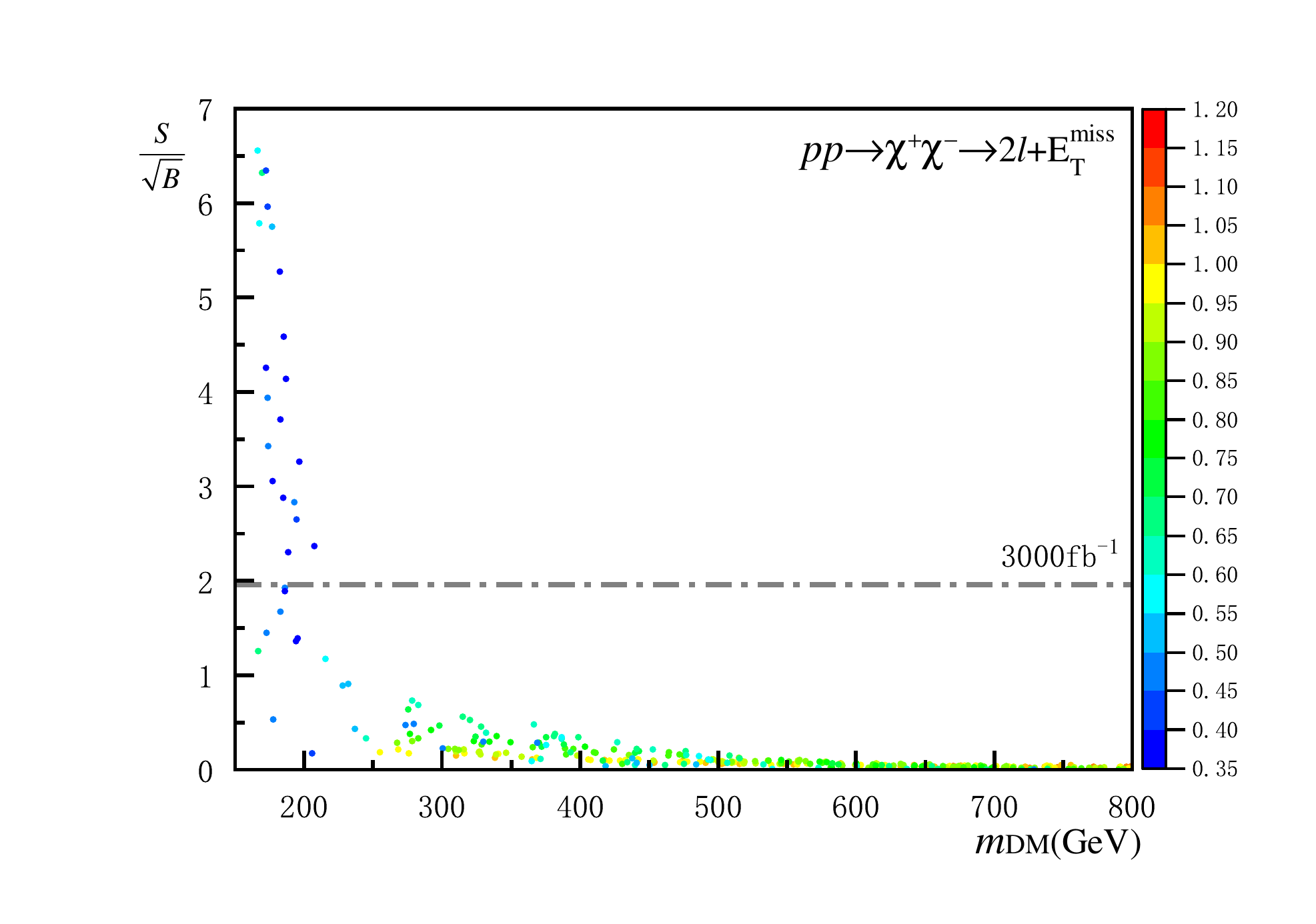}~~~~
\includegraphics[width=8cm,height=7cm]{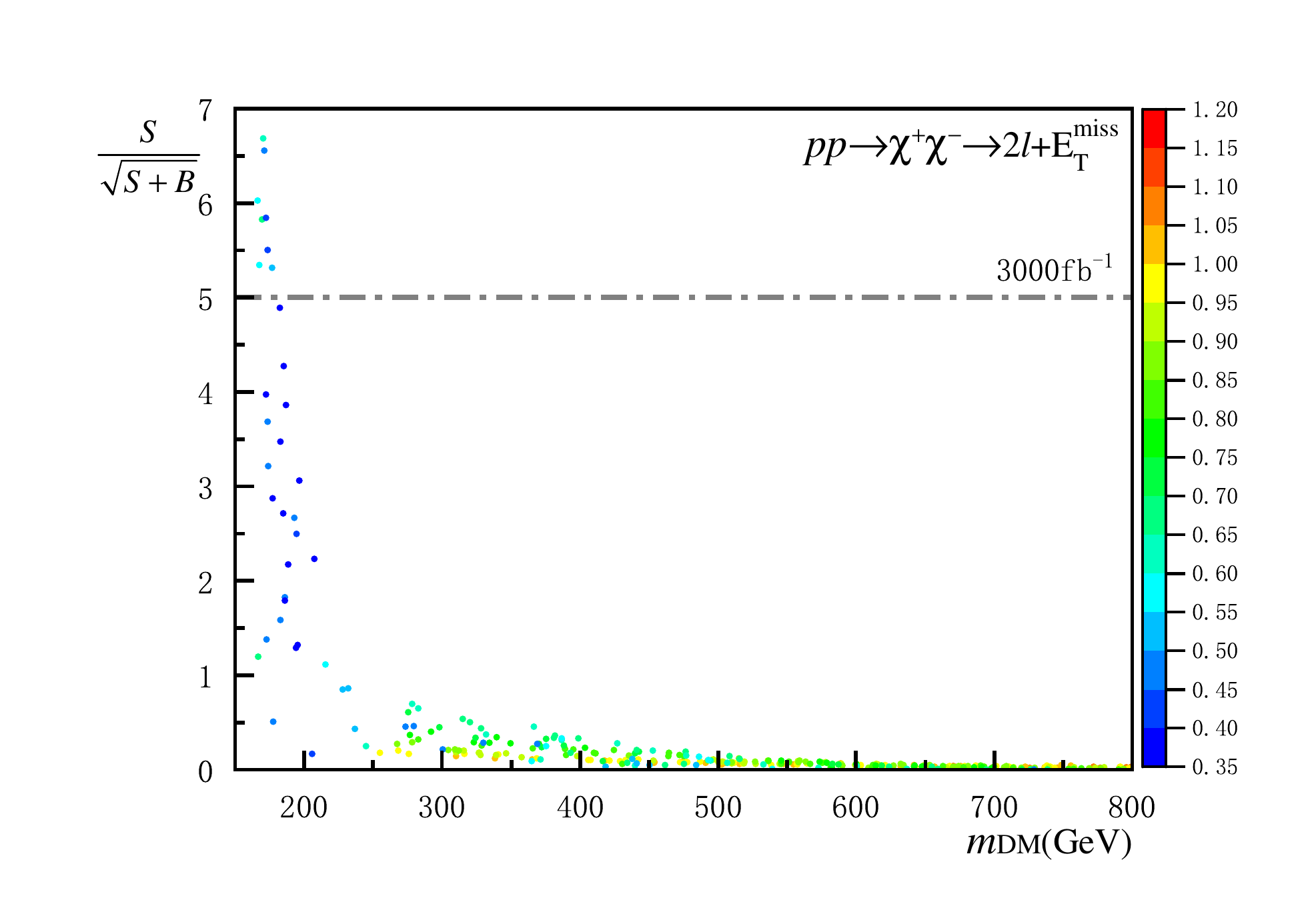}\\
\includegraphics[width=8cm,height=7cm]{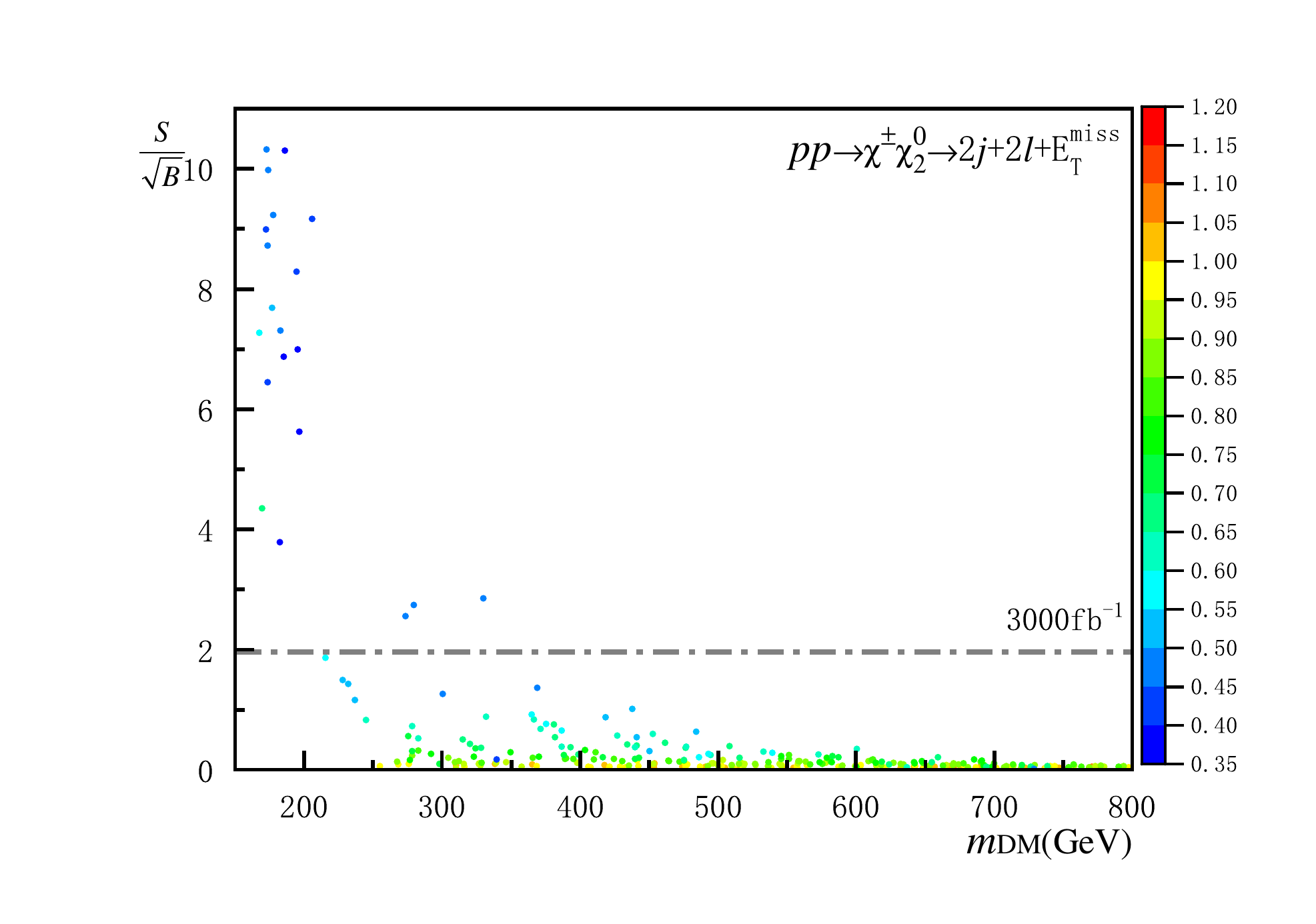}~~~~
\includegraphics[width=8cm,height=7cm]{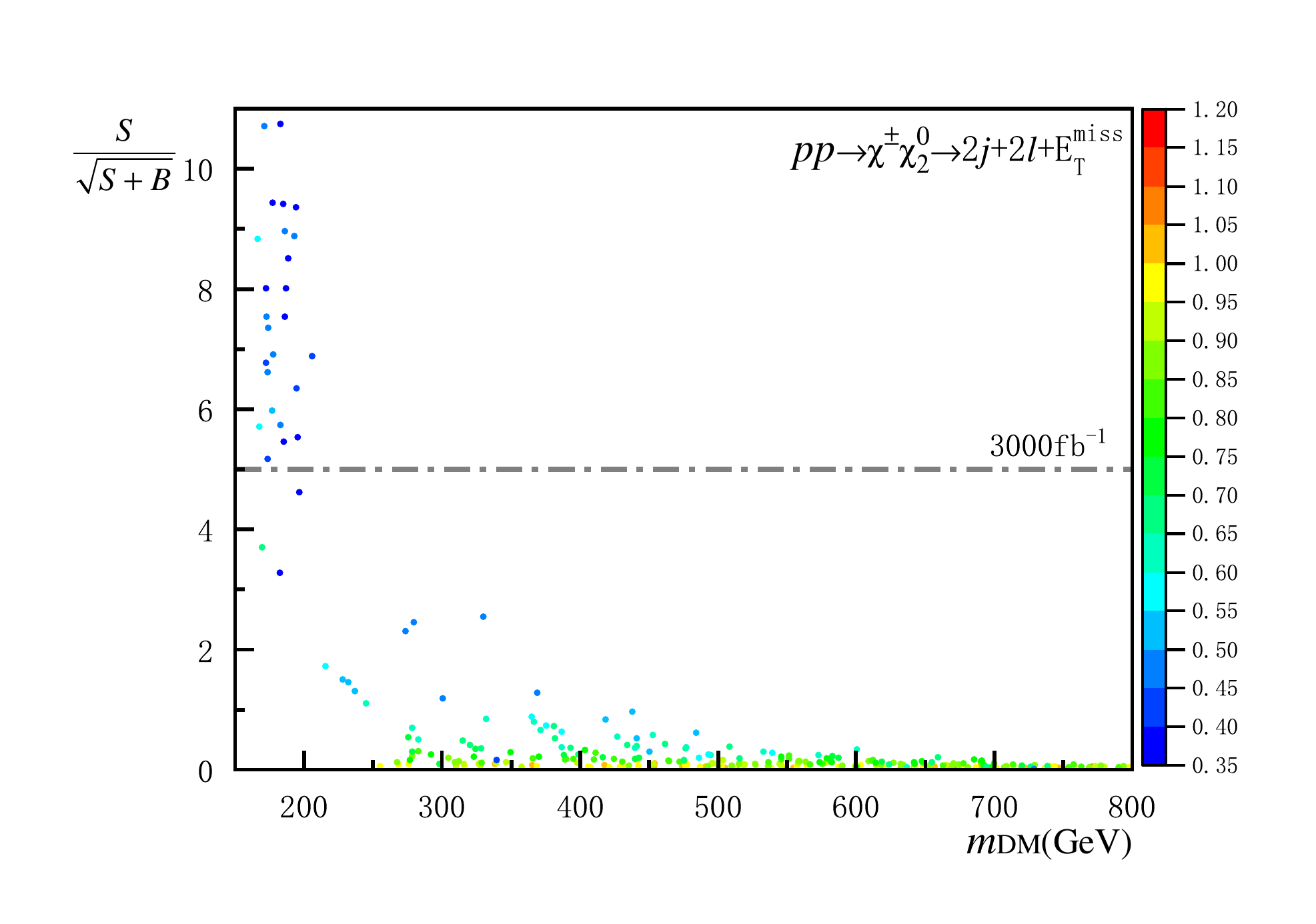}\\
\includegraphics[width=8cm,height=7cm]{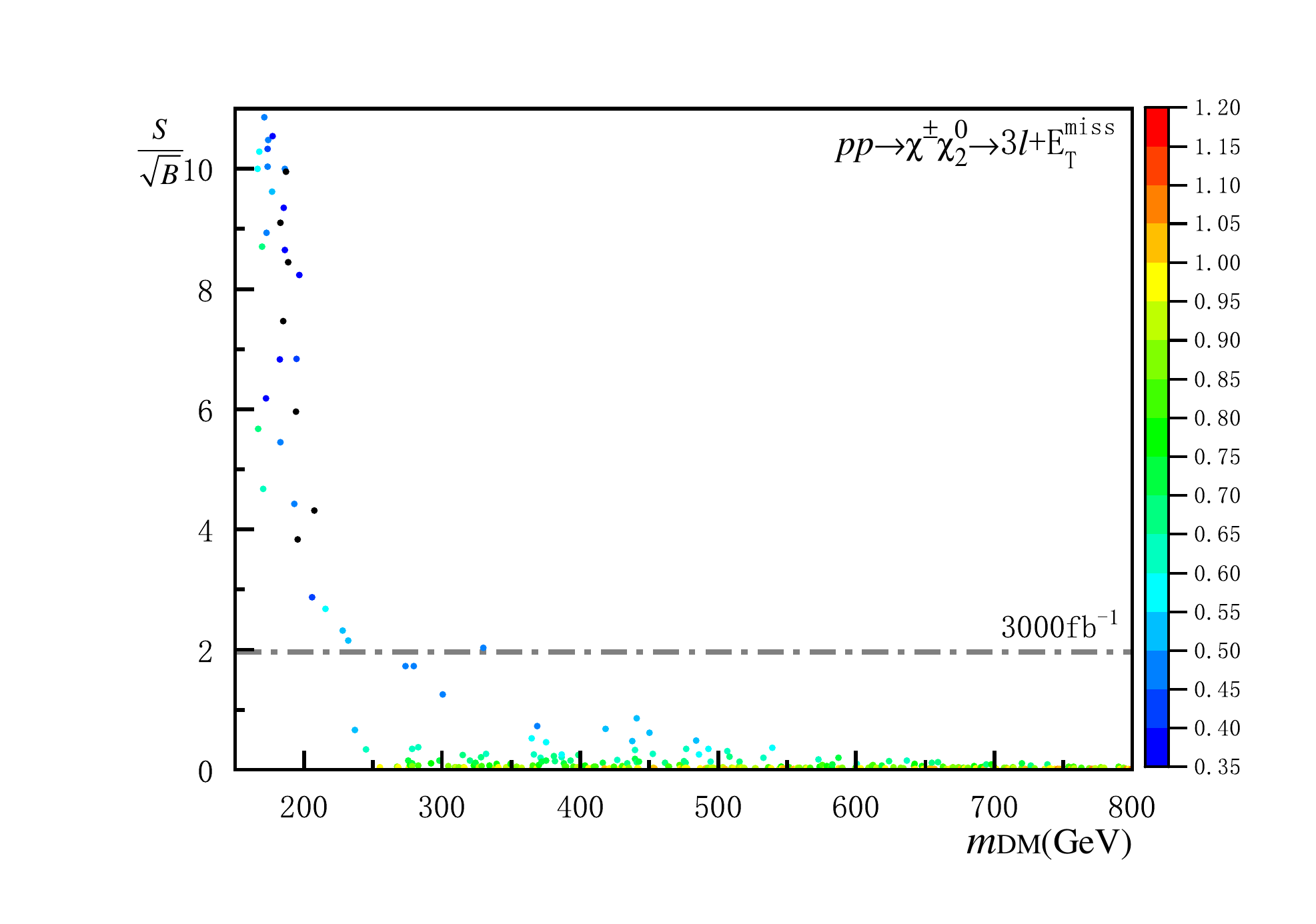}~~~~
\includegraphics[width=8cm,height=7cm]{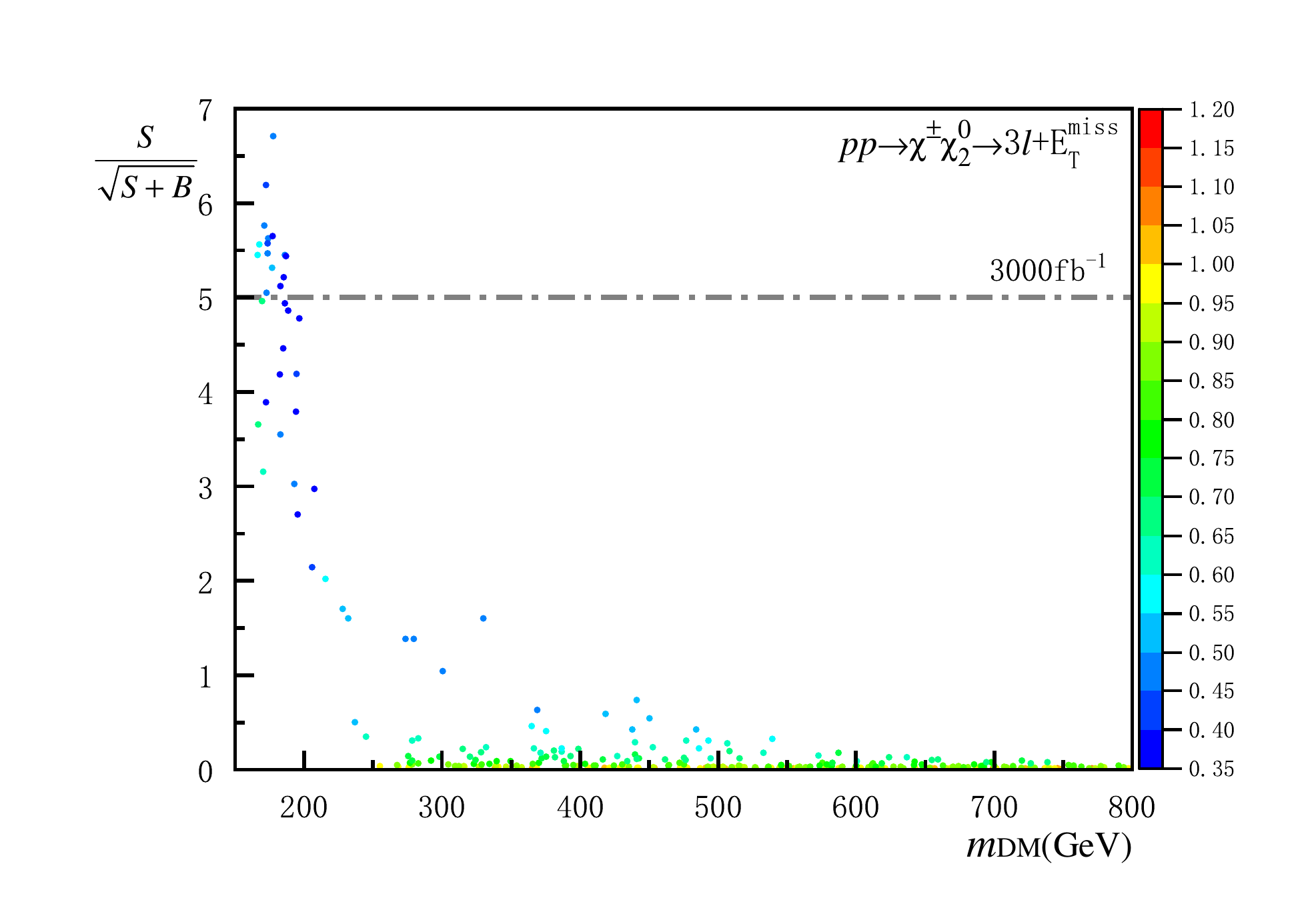}
\caption{The same as Fig.\ref{potential1} with $L=3000$ fb$^{-1}$ instead.}
\label{potential2}
\end{figure*}

\begin{table}
\begin{center}
\begin{tabular}{cccc}
\hline\hline
 $\rm{channel}$ & ~~$\sigma_{\rm{LO}}\rm(fb)$~~  & $N_{\rm{tot}} $ & $N_{\rm{re}}$ \\ \hline
$pp \to W^{+}W^{-} \to 2\ell + E_T^{\rm miss}$ & ~~$3300$   &$10^{5}$ & $51$ \\
\hline
~~~~$pp \to WZjj \to 3\ell +2j+ E_T^{\rm miss}$ & ~~$540$  &  $10^{5}$ & $7$ \\
~~~$pp \to ZZjj \to 2\ell+ 2j+ E_T^{\rm miss}$ & ~~$100$  &  $10^{5}$ & $12$ \\
\hline
$pp \to WZ \to 3\ell+ E_T^{\rm miss}$ & ~~$800$ &  $10^{6}$ & $10$ \\
\hline \hline
\end{tabular}
\label{background}
\caption{Raw data of the simulations of SM backgrounds in Table \ref{channels},
where $\sigma_{\rm{LO}}$ is the leading-order (LO) cross section,
and the efficiencies on the SM backgrounds after cuts imposed are given by $\epsilon=N_{\rm{re}}/N_{\rm{tot}}$,
with $N_{\rm{tot}}$ and $N_{\rm{re}}$ referring to the numbers of events before and after cuts imposed, respectively.}
\end{center}
\end{table}

In terms of the cuts above, we can analysis the signal significances. 
We firstly show in Table \ref{background} the simulations of events of the SM backgrounds.
Then, we show in Fig.\ref{potential1} and  Fig.\ref{potential2} 
both the 2$\sigma$ exclusion (\emph{left}) and  5$\sigma$ discovery (\emph{right}) at the 14 TeV LHC with the integrated luminosity $L=300$ fb$^{-1}$ and $3000$ fb$^{-1}$, respectively.
In these figures, the top, middle and bottom plots refer to lepton pairs $\chi^{+}\chi^{-}$ and $\chi^{\pm}\chi^{0}_{2}$, respectively,
where the efficiencies among the three channels are comparable.
The combination of individual results yields our final observations:
\begin{itemize}
\item  In the coupling range $y\sim 0.35-0.5$ as referred by the dark blue points, 
which is viable for the bino-higgsino system,
DM mass range between $\sim 170-210$ GeV can be excluded at 2$\sigma$ level with with $L=300$ fb$^{-1}$,
and DM mass range between $\sim 175-205$ GeV can be discovered at 5$\sigma$ level with $L=3000$ fb$^{-1}$.
These limits are comparable with those  \cite{1404.0682} of soft lepton final states,
where the exclusion and discovery limits are $\sim 200-280$ GeV and $\sim 130 $ GeV, respectively. 
Earlier discussions about simplified MSSM at the LHC, see refs.\cite{9706509,1409.6322,1510.03460}. 

\item In the coupling range  $y\sim 0.5-0.65$ as illustrated by the blue points, 
which can be applied to the wino-higgsino system,
DM mass range between $\sim 172-210$ GeV can be excluded at 2$\sigma$ level with $L=300$ fb$^{-1}$,
and DM mass range between $\sim 180-200$ GeV can be discovered at 5$\sigma$ level with $L=3000$ fb$^{-1}$.

\item In the coupling range $y\sim 0.65-1.0$ as referred by the  green and orange points, 
which is viable for SD and VL model, 
all of samples with DM mass above $\sim 250$ GeV give rise to very small significances. 
Thus, they are beyond the reach of LHC.
Earlier discussion on this point,  see ref.\cite{0706.0918}.
\end{itemize}

Our analysis above are subject to both the Monte Carlo (MC) based uncertainties and the systematic uncertainties. For the MC uncertainties, the LO cross sections for both the signals and their SM backgrounds can be enhanced by the higher-order QCD effects \cite{Campbell:1999ah,Campbell:2011bn}. The K-factors $k_{b}$ respect to the WW and WZ channel in Table.\ref{background} are of order $\sim1.67$ and $\sim1.85$, respectively, which imply that for the inferred K-factor $k_{s}$ of order $\sim 1.2-1.3$ for the signals, the significance measures in Fig.\ref{potential1} and  Fig.\ref{potential2} are corrected by a factor of order $\sim6\%$ and $\sim9\%$, respectively. For the systematic uncertainties, even although those in the lepton reconstruction efficiency and the b-tagging efficiency etc are small \cite{1803.02762}, the uncertainties related to the jet energy scale and resolution aren't negligible \cite{TheATLAScollaboration:2015hmp}.

\section{Conclusions}
In this work, we have revisited general electroweak DM through the combined searches of DM direct detections and LHC probe of new lepton pairs which appear together with DM in the new physics of weak scale.
Compared to earlier works in the literature, this analysis is more intuitive.
On the other hand, we have utilized a general framework, 
which can effectively describe three well-known extensions on the electroweak sector that contains both DM and new lepton pairs simultaneously. 

The outcomes are two-fold.
First of all, instead of those separate studies either from DM direct detection or LHC probe,
the combination of the two direct detections allows us to explore the real status of electroweak DM.
Second, the general description enables us to uncover more parameter regions unexplored yet.
Utilizing three $Z$- and $W$-associated electroweak processes at the 14 TeV LHC,
we have illustrated that in the parameter space with the lepton pairs decaying to on-shell $Z$ or $W$: 
$i)$ for $y\sim 0.35-0.5$ DM mass up to $\sim 170-210$ GeV can be excluded at $2\sigma$ level with $L=300$ fb$^{-1}$ and up to $\sim 175-205$ GeV can be discovered at $5\sigma$ level with $L=3000$ fb$^{-1}$;
$ii)$ for $y\sim 0.5-0.65$ DM mass up to $\sim 172-210$ GeV can be excluded at $2\sigma$ level with $L=300$ fb$^{-1}$ and up to $\sim 180-200$ GeV can be discovered at $5\sigma$ level with $L=3000$ fb$^{-1}$; 
$iii)$ for $y\sim 0.65-1.0$ DM mass above $250$ GeV is totally beyond the reach of LHC.

The low reaches on the electroweak DM mass in the situation with large mass splitting, together with similar trends in the small mass splitting,  
suggest that new search strategies are needed in order to examine higher electroweak DM mass ranges.
There are a few directions as follows.
Firstly, the ability of detection on the electroweak DM may be improved in other channels that are rarely considered.
Moreover, compared to the sophisticated data analysis adopted here, 
novel methods such as machine learning may provide alternative views, 
see, e.g. \cite{1709.04464,1712.04793}.
Finally, put aside colliders,
one may carefully consider astrophysical probes of the electroweak DM, 
which may be stringent under certain circumstances.

\begin{acknowledgments}
The authors would like to thank Lorenzo Calibbi, Huayong Han, Yang Zhang and Pengxuan Zhu for  helpful discussions.
This research is supported in part by the National Natural Science Foundation of China with Grant No.11775039,
the Chinese Scholarship Council,
and the Fundamental Research Funds for the Central Universities at Chongqing University with Grant No.cqu2017hbrc1B05.
\end{acknowledgments}

\end{document}